\documentclass[a4paper,orivec]{llncs}

\pdfoutput=1

\usepackage{graphicx}
\usepackage{color}

\usepackage{amsmath}
\usepackage{stmaryrd}
\usepackage{url}

\newlength{\spacelen}
\newcommand{\tabspace}[1]
        {\settowidth{\spacelen}{$#1$}
         \hspace*{\spacelen} }

 \newcommand{\infr}[2]
        {\renewcommand{\arraystretch}{1.5}
        \begin{array}{c}
        #1\\
        \hline
        #2
        \end{array}}

\newcommand{\myif}
        {{\bf if}\,}
\newcommand{\mythen}
        {\,{\bf then}\,}
\newcommand{\myelse}
        {\,{\bf else}\,}
\newcommand{\pu}
        {{\bf put}}
\newcommand{\get}
        {{\bf get}}
\newcommand{\delete}
        {{\bf delete}}
\newcommand{\rexec}
        {{\bf rexec}}
\newcommand{\lexec}
        {{\bf lexec}}
\newcommand{\spawn}
        {{\bf spawn}}
\newcommand{\com}
        {{\bf com}}
\newcommand{\ns}
        {{\bf ns}}
\newcommand{\lo}
        {{l}}
\newcommand{\lop}
        {{l'}}
\newcommand{\ok}
        {{\bf ok}}
\newcommand{\err}
        {{\bf err}}
\newcommand{\phbase}
        {{\bf phbase}}

\newcommand{\ipath}
        {{\bf ipath}}

\newcommand{\session}
        {{\bf session}}

\newcommand{\application}
        {{\bf application}}

\newcommand{\exec}
        {{\bf exec}}

\newcommand{\ses}
        {{\cal S}}

\newcommand{\caln}
        {{\cal N}}
\newcommand{\cale}
        {{\cal E}}
\newcommand{\calp}
        {{\cal P}}
\newcommand{\call}
        {{\cal L}}

 \newcommand{\nil}
        {\underline 0}

 \newcommand{\eqdef}
        {\buildrel \Delta \over =}

\newcommand{\cali}
        {{\cal I}}
\newcommand{\calr}
        {{\cal R}}

\newcommand{\bs}
        {\backslash}
\newcommand{\pa}
        {\mathop{\Vert}\,}
        
\newcommand{\auxarrow}
        {\mathop{\longrightarrow}}
\newcommand{\notauxarrow}
        {\mathop{\not \!\! \longrightarrow}}
         
\newcommand{\arrow}
        {\, \auxarrow\,}
\newcommand{\parrow}
        {\, \auxarrow_{c} \,}
\newcommand{\nparrow}
        {\, \notauxarrow_{c} \,}

\title{File Managing and Program Execution in Web Operating Systems\thanks{Research partially funded by 
       EU Integrated Project Sensoria, contract n. 016004.}}
\author{Mario Bravetti}
\institute{Department of Computer Science, University of Bologna, Italy\\
           E-mail: {\tt bravetti@cs.unibo.it}}

\date{}

\begin{document}

\maketitle

\begin{abstract}
\noindent
Web Operating Systems can be seen as an extension of traditional Operating Systems where the addresses used to manage files and execute programs (via the basic load/execution mechanism) are extended from local filesystem path-names to URLs. A first consequence is that, similarly as in traditional web technologies, executing a program at a given URL, can be done in two modalities: either the execution is performed client-side at the invoking machine (and relative URL addressing in the executed program set to refer to the invoked URL) or it is performed server-side at the machine addressed by the invoked URL (as, e.g., for a web service). Moreover in this context, user identification for access to programs and files and workflow-based composition of service programs is naturally based on token/session-like mechanisms. We propose a middleware based on client-server protocols and on a set primitives, for managing files/resources and executing programs (in the form of client-side/server-side components/services) in Web Operating Systems. We formally define the semantics of such middleware via a process algebraic approach.
\end{abstract}

\section{Introduction}

The widespread use of more and more powerful mobile devices, like smartphones, in addition to personal laptops, laptops
used at work, etc... has led to the need of exploiting the Internet as a repository for storying personal information
and applications/programs so to be able to use them from 
any of these devices, not to loose them in the case 
one of these devices is destroyed/stolen and not have to re-install/re-configure them when such personal devices are
changed: personal cellphones/smartphones tend, e.g., to be changed much more frequently than laptops. These needs have led to the recent development of CLOUD computing which shifts all resource managing from local machines to a remote (set of) server(s) located somewhere in the Internet. 
Such a trend is, however, influencing much more deeply the 
way in which people use personal computers: browsers are, by far, the most used computer application and play, more and more,
the role of operative systems, which allow the user to use
application/programs and retrieve/store information. Another
reason of this trend is related to the capability of (web) applications and information deployed in the Internet to be shared among several users, thus allowing for cooperation and
enhanced communication. Examples are Google web applications (Google docs) and social networks like Facebook. 

Being the computing experience and the evolution of computer languages/technology more and more related to just the browser,
this naturally leads to the idea of: from the one hand making its functionalities to become part of the operating system, from
the other hand getting free from the more traditional (and heavy) way of installing and configuring applications. 
In essence, the shift from traditional operating systems to
so-called Web Operating Systems consists in changing from 
usage of local filesystem path-names to manage files and execute programs (via the basic load/execution mechanism)
to usage of URLs. A first consequence is that, similarly as in traditional web technologies, executing a program at a given URL, can be done in two modalities: either the execution is performed client-side at the invoking machine (and relative URL addressing in the executed program set to refer to the invoked URL) or it is performed server-side at the machine addressed by the invoked URL (as, e.g., for a web service).
From the viewpoint of application development and execution,
WEB-OS allows applications to be deployed anywhere in the Internet
and used from any machine by exploiting a front-end/back-end philosophy typical of web applications. 
In WEB-OSes a typical application will have a front-end consisting of several {\it application components}, i.e. a (graphical) user interface, which is executed client-side and a back-end which consists of several {\it service components} remotely executed on the machine where the application is deployed (such a back-end may in turn exploit other resources like databases leading to the typical multi-tier architecture of classical web technologies). Notice that, thanks to the usage of relative URLs in {\it application components}, which
are resolved relatively to the romote directory URL from which  the component has been downloaded, they can access remote
resources/service components in their delployment environment independently from the location of their execution environment. For example a typical WEB-OS application has two basic forms
of file reading/saving: relative to the machine where it is deployed (default way of resolving relative addresses) or absolute/relative to the machine where the user is using it.

The aim of this paper is to propose an architecture and a design/implementation for a WEB-OS which uses the mechanisms
above as the ``normal'' way for executing and deploying 
programs (to be used by the local machine or by other machines
over the Internet). The idea has been to perform, apart from
taking some concepts and basic mechanisms from traditional
web technologies, a complete fresh re-start in developing
such a WEB-OS architecture. In particular, while mechanisms
typical of a browser (and a server) are present, we abandon
the usage of HTML as the ``central'' format (often also a bottleneck) for user-interfacing at the client-side, around
which the client-side technologies must operate. Notice that
presence of HTML in browsers is one of the main reason that
led to the current success of interpretation/javascript based
client-side web technologies over the virtual machine based ones
like, e.g., java applets.
In a WEB-OS, where execution/development of web applications, 
becomes the ``normal'' way of executing/developing applications,
we believe that interpreted technologies and HTML are no longer
adequate as the only user-interface technologies. Front-end
of applications are more naturally implemented by virtual-machine based graphical user interfaces, so to have
interfaces with the same quality as the traditional standalone
OSes applications. Such a belief is also along the lines of current technology trends: application development in smartphones (which use graphical interfaces) and development of rich Internet application development where execution outside the browser is often considered a preferred feature (e.g. in Java by using the Java Web Start feature).

A confirmation that the current trend is bringing towards the
development of WEB-OSes and, more in general, platforms 
for executing installation-free applications deployed in the Internet which remotely preserve configuration and data, is given by the very recent presentation
of the Google Chromium OS~\cite{COS}, which has been developed concurrently with the ideas/implementations presented in this paper. From the available information it seems however that
such an OS does not follow the ``radical'' approach taken in this paper and somehow resambles browser functioning also
on usage of HTML pages. A consequence is that it considers usage of interpreted/javascript technologies as the client-side
application front-ends (as in Google docs).

The WEB-OS architecture presented in this paper, while being freely designed from skratch, allows for any current client-side/server-side web technology/languages 
to be used in developing front-end/back-end of programs
(both interpreted and virtual-machine based for the front-end,
of arbitrary nature for the back-end)
and any possible format for data exchanged (XML, Java object streaming, Javascript JSON, etc...).
The idea is to have a language/technology and data format independent middleware and an extensible set of client/server plug-ins one for each of the supported technologies, which are used to interface the middleware to application/service components developed with such a technology.
We propose a model of such a middleware based on: a set of primitives for managing files and executing programs in WEB-OSes; and mechanisms for application/service component (un)deployment and execution.

In such a middleware, an important role is played by session
managing as a mean for user authentication and for the realization of complex workflow/business process patterns~\cite{BPEL}. To this end we will also consider session
delegation as a basic mechanism of the WEB-OS middleware.

After presenting the basic architecture of the middleware we will propose a service-based implementation for it which uniformly combines file/resource managing with service component remote execution. We will do this by developing an extension (both at the conceptual and at the technical level) of restful Web-Services~\cite{REST} which 
use HTTP and its request methods as the communication
protocol. The adequateness of this implementation is confirmed
by the current success and widespread use of restful Web-Services and of HTTP POST for interface-based
service components combined with session-like mechanisms
(based on session id or authentication token). Facebook and Twitter are, e.g., endowed with a wide set of services that
are in this form.

In order to unambiguously present the behaviour/semantics of such a middleware and of applications it executes,
we use a process algebraic~\cite{PICALC,SCC,COWS} approach to
formally define the semantics of the middleware primitives and component deployment/execution. The process algebra that we present is, to the best of our knowledge, the first one representing (via URL managing) the behaviour of restful Web Services (furthermore it extends such behaviour) and the first one managing (via primitives which do not deal explicitly with session identifiers as data) sessions in the form of pairs: session identifier and context/application it refers to.

The paper is structured as follows. In Sect. 2 we present
the basic WEB-OS architecture, in Sect. 3 we present the 
service-based implementation, in Sect. 4 we present 
the formalization with the process algebra.
In Sect. 5 we make some concluding remark.

\section{Basic Architecture}

\begin{figure*}[t]
\hspace*{-2.8cm}
\includegraphics[width=17.5cm]{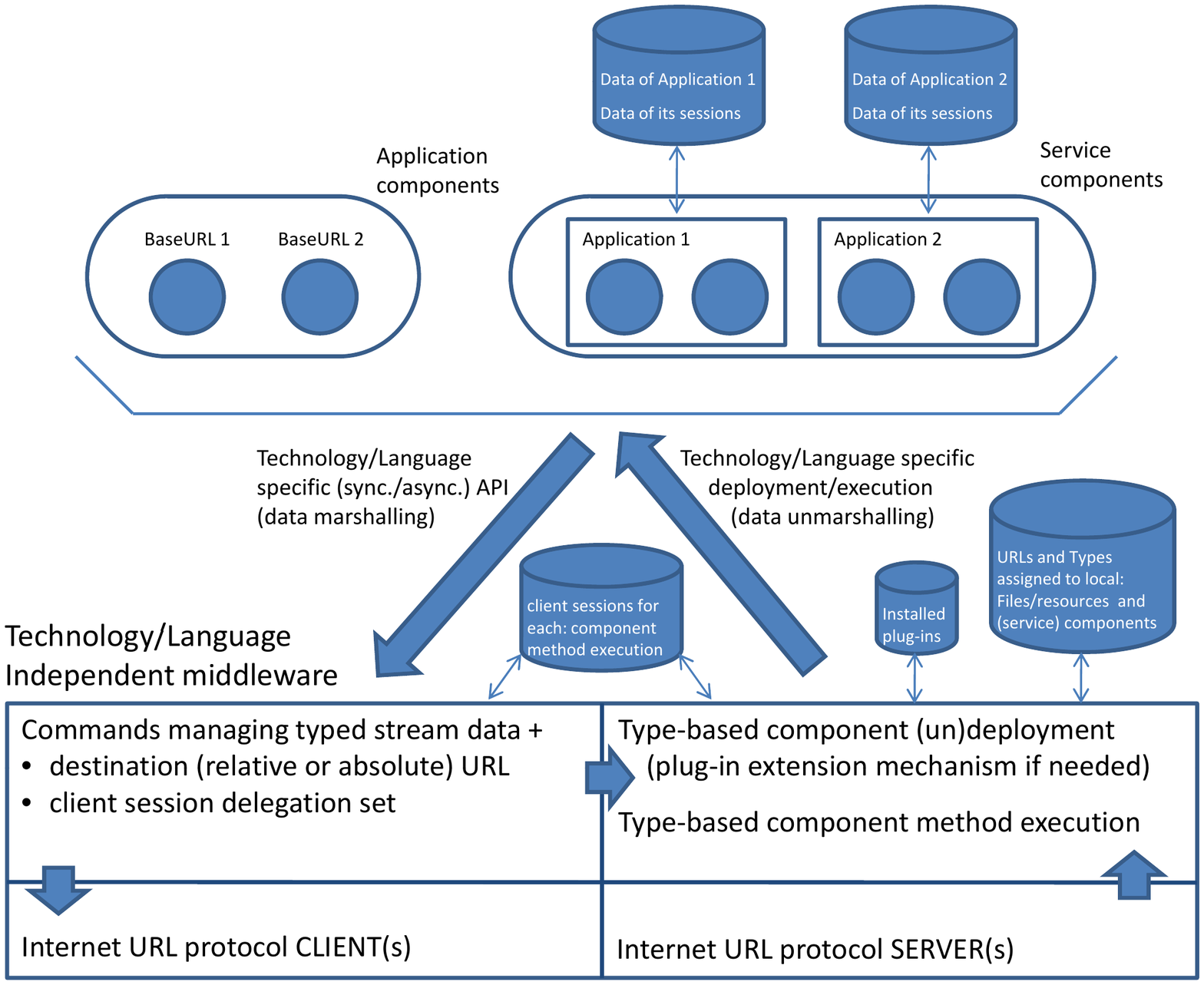}
\caption{Basic WEB-OS architecture}
\label{figarc}
\end{figure*}

We now need to introduce some terminology on URLs that we will use to present the basic architecture of the WEB-OS.

URLs can be of two kind: URLs ending with ``$\backslash$'',~\footnote{For denotational convenience, in this paper we use backslashes in URLs instaed of slashes, because we use slashes to represent replacements.}
 representing resource collections (directories) and URLs ending
with an identifier, representing a single resource (e.g. a file).
A URL of the first kind can be used as a ``{\it base URL}'' that includes
several resources whose ``URLs'' have such a base URL as a syntactical prefix. A particular case of base URL is 
a ``{\it context URL}'' that is used to denote
all the resources and services belonging to an entire (web) Application. We will also use the intuitive (even if a little improper) terminology of ``{\it relative URL}'' as: a pathname (an alternated sequence of names and ``$\backslash$'' characters) starting with a ``$\backslash$'' or not. In the former case the relative URL is called, more specifically, a ``{\it root relative URL}'': when the
former is {\it resolved against a base URL}, we get an absolute (i.e. non-relative) URL by replacing the (non-context) pathname part of the base URL
with that of the root relative URL; when, instead, the latter is resolved against a base URL, we get an absolute URL by just concatenating the two (or by performing the preliminar name
eliminations in the case ``$..$'' are used).
These notions will be formalized in Section 4.

Another important notion that we will use is that of session.
We will assume that each (web) Application uses sessions
(session identifiers) to maintain data associated with client 
sessions via a technology dependent data structure (e.g. for Java based technologies, session attributes containing objects). As quite usual, we will consider each Application
(identified by a context URL) to independently manage sessions,
which, hence, have application scope. From the client side, therefore, session information will be collected as set of pairs 
composed of a {\it session id} and the {\it context URL} the id
refers to. 

Figure~\ref{figarc} shows the WEB-OS architecture. 
The upper part of the figure shows deployed/active, i.e. under execution or waiting for method calls, application and service components (assuming that application components, that interact with the user by, e.g., some graphical user interface, have been installed by downloading them from some ``Base URL'').

Concerning the lower part,
the language/technology independent middleware receives (via API) 
request to execute commands from application and service components. 
All such commands (that we will list in the following)
receive as argument a (possibly empty) typed data stream (file) and return a (possibly empty) typed data stream (file). The file
type determine the format of its content and can be implemented, e.g., by using the mime-type standard. Additional arguments
of such commands are the command destination (absolute or relative) URL and session delegation set. The latter is a set of {\it context URLs} whose
client sessions (if detained by the component execution calling the command) are to be passed, besides the client session
related to the destination URL (standard session managing), to a component which is, possibly, executed as the effect of the command. As we will see, the middleware maintains a set of detained client sessions (pairs session id and related context URL) for each execution of an (application or service) component. Client sessions are therefore not shared between
components or between different execution of methods in the same service component, differently from what happens, e.g., in a browser.

Relative addresses are resolved differently depending on the
invoking component: If it is an application component than
the address is resolved against the code-base URL (i.e. the URL
of the directory the application component was downloaded from)
which is stored together with the application component when
it is locally installed (denoted by ``Base URL'' in the figure). If, instead, it is a service component
then the address is resolved against the physical-base URL i.e.
the directory containing the URL used to execute the service component.

Notice that the technology/language dependent API, which invokes middleware commands, does not need to wait to receive the return value from the middleware before continuing with the execution of code. It can install an event listener (as done e.g. in AJAX)
and interface with the middleware in such a way that the listener is executed upon completion of the command. Independently of the
synchronous or asynchronous realization of the API for a particular technology, the middleware must be implemented so to be able to manage command requests in parallel in that:
the client technology may use multithreading and, anyway, middleware command requests may come from method executions of (different) components, which are run in parallel.

Middleware basic commands (managing typed stream data) are the following ones:
\begin{itemize}
\item {\bf Read} (inputstream), {\bf write} (outputstream) or {\bf delete} file/resources with the specified URL: writing
creates or modifies and has a typed file as its parameter, while a successful reading returns a typed file.
\item {\bf Remotely execute} an operation of a service component with the specified URL: it takes a typed file as the
parameter to be passed to the operation and returns a typed file as the operation result.
\item {\bf Deploy} or {\bf undeploy} service components
at the specified local URL:
deployment takes a typed file containing the component code and, if needed, deployment infos as the parameter. Undeployment returns
a typed file in the same form, so to allow for component migration.
\item {\bf Locally execute} an application component by reading the file with the specified URL which contains its code: it takes a typed file containing the parameter to be passed to the component execution/initialization and, if needed, deployment infos and
it returns a typed file containing a local reference to the deployed component.
\end{itemize}
Optionally, the architecture could also include commands
for execution of application component methods
(besides direct interaction of the user with the application component user interface).
In this case the reference returned by the local execution
command (and which is then used as a destination address to execute methods) must be a ``typed'' reference, i.e. a reference which makes it possible to determine the type of the component. 
Moreover, deployment/undeployment commands could be extended to
work on any URL (and not just local ones) so to realize romote
mechanisms typical, e.g., of a CLOUD system. However it is maybe more realistic to think about romete (un)deployment of components
as a higher level mechanism that is not directly implemented by the middleware via a unique command and that could be technology
dependent in the use of the related Internet protocol(s)
(it could involve invocation of serveral middleware commands in
a technology dependent way).

Command execution is based on direct usage (as a client) of the protocol specified in the URL address and/or, in the case of service components (un)deployment and application component
local execution (and, if considered, of subsequent calls to its methods), of the middleware functionalities presented 
in the right-hand side of figure~\ref{figarc}.

Concerning client URL protocol invocation, the command
performs a request containing the typed file parameter, if any, and (by analyzing the client sessions detained by the method that invoked the command):
the session id of the client session (if any) associated
with the destination URL and the session ids of the client sessions (if any) associated with the session delegation URL contexts (the latter information is passed as pairs session id and related context URL).
The response, symmetrically, is expected to include, besides 
a possible typed file or error message, a
possible
new status of the client session related to the destination
URL (as standard) and of the delegated client sessions (which
are delegated ``back'' at the response). The former is returned
to the caller of the command, while the latter is used to update
the status of the set of client sessions detained by the method that invoked the command.

Concerning the right-hand side of figure~\ref{figarc},
service component (un)deployment is based on the type of the
component to be (un)deployed (indicating the technology/language of the component): in the case of deployment, if the service component technology is still not supported by the WEB OS running on the local machine,
a remote repository of plug-ins is queried to find a plug-in to
install in the system so to support the service component technology. For supported technologies it 
uses the corresponding plug-in to perform the (un)deployment
and it associates the deployment URL (in the configuration of the server of the protocol specified in such URL) and the type with the component.
Similarly, application component local execution is performed
by doing, at first, a deployment of the application component 
based on the type of the
component to be deployed: as for service components, if the application component technology is still not supported by the WEB OS running on the local machine,
a remote repository of plug-ins is queried to find a plug-in to
install. Once the application component is deployed by using 
the plug-in corresponding to its type, the component is executed/initialized by exploiting the same plug-in.
Before actual application component execution, the middleware
associates a set of detained client sessions to such an execution
(which is initialized with standard and delegated client sessions
passed with the command) and maintains it until the end of the execution.
If subsequent invocations of the application component methods
are possible, the typed reference identifying the component
will cause again the same plug-in to be exploited.

As we already mentioned, the WEB-OS middleware is also endowed with server(s) for protocols of URL addresses of files/resources and service components it exposes to the Internet. According to the client invocations we described above, such server(s) receive requests to read, write, delete files/resources and to remotely execute a method of a service component. Concerning files/resources, we just assume that such
server(s) perform the required operation by exploiting standard OS functionalities (e.g. of the underlying local filesystem manager), and, whenever a new file/resource is created the server records its URL and its type. The client session passed with the
request can be exploited to check the identity of the client
and its right to manage the file/resources.
Concerning service components, the server exploits the middleware functionalities to execute the required method
via the plug-in corresponding to the component type.
Similarly as for application components, before actual method execution, the middleware
associates a set of detained client sessions to such a method execution (which is initialized with standard and delegated client sessions received with the request) and maintains it until the end of the execution.

Notice that, in the case middleware commands are used that refer to local URLs, there is no need to actually perform
a protocol client invocation and then receive the request from the corresponding local server: in the case the addressed URL
is local we can avoid invoking the protocol client and, instead, directly perform operations on the local resources/filesystem and/or invoke directly the right-hand middleware functionalities.

Finally, concerning the interface of the middleware with the
technology/language dependent part of the WEB-OS, which is realized by an extensible set of service or application technology plug-ins:
\begin{itemize}
\item The (service or application) component code invokes the
middleware commands via a technology/language dependent API
which also performs data marshalling.
\item The middleware executes (methods of) service or application components by performing technology/language dependent required mechanisms and also performing data unmarshalling.
\end{itemize}
The type of file used for data marshalling (e.g. XML or JSON for Javascript) and unmarshlling depends on the plug-in
associated with the component type and on its configuration.

\section{Service-based Implementation}

We propose a service-based implementation for the basic architecture presented in the previous section.
In particular the implementation that we propose aims at adopting a uniform mechanism for managing both file/resources and service
components. In the following we will call ``operation'' a method
of a service component.

The needed mechanisms for dealing with files/resources are provided
by so-called restful Web Services~\cite{REST}.
In Restful Web Services the HTTP protocol and HTTP request methods GET,PUT and DELETE 
are used to manage with the resource
identified by the URL address of the request.

In the most simple case the URL corresponds to a local directory/file
in the destination machine (e.g. according to a mapping from the context part of the URL to a directory in the machine filesystem)
and the invoked HTTP server performs the operation corresponding to the request method: read, write (create or modify), delete.

However, in restful Web Services, URL addresses can be used to represent resources different from directory/files, e.g. to directly manage records of a database table. The idea is to manage addressed
resources via a service component which is associated to a {\it pattern} in the form ``$\backslash pathname \backslash *$'': all requests to URLs whose (non-context) pathname matches the pattern
(where ``$*$'' stands for any possible suffix) cause an execution
of the associated service component which possesses an operation for each request method (for simplicity we assume the executed operation to have the same name of the corresponding request method).
The operation receives the ``suffix'', that in the invoked URL
replaces ``$*$'', as a parameter and uses it as an identifier for the resource (e.g. a database record) on which it actually operates
by executing resource specific access code.
Notice that resource URLs managed by a service component can be
of two kind: the resource collection kind (the URL ends with ``$\backslash$'') and the single resource kind (the URL ends with a name). On an URL of the former kind it is possible to also use the
POST HTTP request method with the following intended behaviour: a subresource (a resource whose address is ``$URL \; name$'' or ``$URL \; name \backslash$'' for some ``$name$'') is created whose content is passed as the body of the request (as for PUT) and whose name is a ``new'' (fresh) name established by the server.

Notice that, in restful Web Services, it is not mandatory for
service components to manage several resources: the pattern 
``$\backslash pathname \backslash name$'' (or just ``$\backslash name$'') can alternatively be used, which matches just a single resource. In this case operations of the service component do not receive a resource identifier, but just manage the single resource represented by the only matching URL.

The interpretation of service components as manager of resources
adopted in restful Web Services leads to service components
endowed with an operation for each HTTP method. This approach
can be smoothly extended so to encompass also interface-based service components, i.e. those of the kind we considered in the previous section, which: possess any number of operations with arbitrary names, are invoked by specifying the name of the operation and do not possess a pre-defined intended semantics. Moreover, this can be done by preserving the interpretation of service components as manager of resources. To better understand this, we make a parallel with object oriented programming. 
The resources managed by a service component can
be seen as objects of a class (in this case the resource is the memory occupied by object's fields) and service component operations as methods defined in the class. In this view, URL suffix identifying the resource plays the role of the object reference which is passed to service component operations similarly as the
reference ``this'' is passed to class methods. Moreover operations
with the name (and meaning) of the HTTP request methods correspond to
getter,
putter,
constructor 
and destructor methods, 
i.e. methods which in a class/component have a special pre-defined meaning. The presence of such special (related to object resource managing) methods in object classes, does not prevent them to also
possess (non static) method with arbitrary names which act on the object resources in an arbitrarily-defined manner. 

Technically, such user defined operations can be added to the 
restful Web Service paradigm by exploiting POST requests on 
single resource URLs: the body of the request contains the invoked
operation name and the parameter typed file; the body of the 
response the typed file with the return value.
By following this approach and by continuing the parallel with object oriented programming we have that:
\begin{itemize}
\item Statics of a class correspond to collection resources managed by a service component, which besides being themselves managed as resources have the ability to construct new subresources via POST (as for the static constructor method) and return a reference to them.
\item Non-statics of a class correspond to single resources managed by a service component, which 
besides being themselves managed as resources can be manipulated 
by arbitrarily-defined operations invoked via POST. 
\end{itemize} 

Summing up, the service-based implementation is based on the HTTP protocol. The middleware commands described in the previous section
are implemented as follows. We have GET, PUT, DELETE commands which cause HTTP client requests with the corresponding method to be performed and the REXEC (remote execution) command that cause a POST HTTP client request to be performed (and the parameter file sent as the request body). LEXEC command retrieves the typed file with
the application component with a GET request.
From the HTTP server side, when a request is received, the destination URL is checked for matching with patterns of deployed components and, in case of successful matching, the longest
matching pattern is selected (deployment of multiple components with the same pattern is not allowed) and the requested operation (that 
corresponding to the HTTP request method) of
the associated service component executed.~\footnote{This is analogous to the mechanism for determining the servlet to be executed in a Tomcat server.} 

Notice that, the extension of restful WS we are considering, which allows arbitrarily-defined operations to be invoked via POST, makes
it meaningful to use sessions and session delegation. This because,
while in restful WS~\cite{REST}, since the only allowed operations
on resources are construction, destruction and getter, putter methods (i.e. those that also in object oriented programming are ``special'' methods with a predefined intended semantics and which
provide the fundamental mechanisms to manage resources in terms of object memory), it does not makes sense to make their behaviour/semantics to depend on sessions (i.e. on data recorded by previous operation invocations by the same client); in our extended setting, instead, since we are considering also arbitrarily defined additional methods, it is correct to have their semantics to possibly depend on data recorded by previous invocations.

We guarantee that service components possess an operation for each request method by defining
default behaviours (which are performed in the case no operation is explicitly defined). In the case of GET, PUT, DELETE requests the
default behaviour is to perform the corresponding action on the local filesystem at the corresponding location (according to a mapping of the URL context prefix into the local filesystem)
and ignoring session information, including delegation, passed with the request. Concerning a POST request the default behaviour of the invoked component REXEC operation is the following.
If the destination URL is a resource collection it generates a typed file (or directory) with a fresh name in the local filesystem directory corresponding to the destination URL containing the body of the request and it responds with the generated name (ignoring session information, including delegation, passed with the request); if, instead, it is a single resource, it expects
the body to be a combined file (e.g. a combined mime-type) containing an operation name ``$op$'' and a parameter file, it invokes operation ``$op$'' of the service component and it responds with the returned value.

When defining GET, PUT, DELETE, REXEC operations in service components it is possible to invoke default versions of 
commands and explicitly pass the URL suffix to them, that identifies the resource they must manage. A typical use of this pattern
is to use the received session to check the identity of the client
and then to invoke the default behaviour.

Notice that, it also makes perfectly sense to add arbitrary
defined operations, besides construction af a new subresource, also to collection resources (which would correspond to static methods different from the constructor in object oriented programming). This is realizable by just modifying the default behavior of REXEC so to manage combined ``$op$''-parameter file
requests also on collection resources.

Concerning service component deployment and undeployment commands, they must now associate a URL pattern (and not just a single URL) to deployed component. 

Some considerations are now in order: the adoption of
a resource based approach (originating from restful Web Services) allows us to manage general resources via URL (e.g. databases) 
and not just files. This is made possible by association
of patterns to service component identifying the resources that it manages, i.e. service components are not identified by URLs
which instead {\it just represent resources}: they are invoked by performing an action on one of the resources that they manage. In the case of a single resource pattern the action must be performed on the unique resource that they manage and, only in this case, we have a simple scenario where they are accessed by a unique URL.
Effectively dealing with general resources is also made possible
by the restful Web Service mechanism of creation of fresh subresources, i.e. the intended behaviour of HTTP POST on collection resources.
The extension we propose of usage of HTTP POST (REXEC) on single resources makes it possible to integrate resource managing and service component operation invocation (that looked like being dealt with separately in the architecture of the previous section)
in a uniform resource-based mechanism.
Finally, the capability of managing client session and client session delegation (introduced in the previous section) makes 
service component capable to implement complex workflows/businness
processes. On the other hand practical applications (like facebook and twitter services) have shown the importance of user authentication/session managing via access tokens/session ids.

\section{Formal Representation of Middleware Behaviour}

We will now present the process algebra formalization for 
WEB-OS application behaviour which encompasses formalization
of the middleware primitives and API invocation mechanisms.

In order to be able to cope with all the mechanisms we will make
the following assumptions that allow for a more uniform treatment:
\begin{itemize}
\item Application components are deployed in the same way as service components: their reference is single pattern URL that can be used to invoke their methods via REXEC. Such methods can be used to abstractly represent interaction of the user with the components.
\item Deployment/undeployment is performed by simple PUT and DELETE commands at special (local or non-local) URLs dedicated to delpoyment of application and service components (that act as manager of the resources located at the associated URL pattern). GET can be used for component migration.
\item Commands in the process algebra are prefixes that wait for completion before performing the continuation code. This is meant to represent the behaviour of commands at the middleware level.
If we are representing a technology/language with an asynchronous API we need to explicitly express in the process algebraic specification the asynchronous behaviour of the API with: multithreading and communication related to notification of command completion.
\end{itemize}

\subsection{Syntax}

We use $x, y,\dots$ to denote generic names (identifiers) over a set $\caln$. Moreover we use
$\lo,\lop$ to denote locations overa set $\call$.
Locations represent application contexts, i.e. a location $l$ 
identifies both a server (IP address + port) and one of its application contexts.

The process algebra is an extension of pi-calculus~\cite{PICALC} that represents the Internet as a network $N$ of 
resources $R$ deployed at some URL $url$: ``$[R]_{url}$''.

$$N ::= [R]_{url} \mid N \pa N \mid (\nu x) N$$

The restriction $(\nu x)$ is used as in pi-calculus to represent 
the scope of a name (i.e. encloses the program/resources that have access to it).

URLS $url$ are defined as pathnames starting with a context location $\lo$. We consider a special context directory called
$\exec$ that we will use to deploy components, i.e. their code
and their associated info, as the URL pattern they manage.

$$ url ::= burl \bs x \mid burl \bs $$
$$ burl ::= \lo \mid \lo \bs \exec \mid burl \bs x$$

We now also introduce relative pathes $rpath$ that will be used in the following, e.g., to represent ``URL suffixes'' which identify a resource belonging to a given pattern

$$rpath ::= bpath \mid bpath \; x$$
$$bpath ::= \varepsilon \mid bpath \; x \bs$$

and patterns $pat$ , that will be associated to components

$$pat ::= \bs bpath \; x \mid \bs bpath \; x \bs *$$

Resources $R$ can be either values $v$ (typed files) or programs
under execution $P$.
$$R ::= P \mid v$$

We will see the syntax of programs $P$ in the subsection about semantics. Concerning values $v$ we consider primitive values $pval$ which should at least allow us to represent successful
and erroneous response from a request an numbers, names $x$,
passive typed (where the type denotes their technology) application components $\langle D \rangle_{type}$ to be downloaded by LEXEC,
deployed components $\langle D \rangle^{url^{\bot} \rightarrow pat}_{type}$,
pairs $x<v>$ used to represent combination of operation/method name $x$ and parameter value $x$ passed to a REXEC, references
$ref$ that we will define in the following as URLs or relative pathes, etc..

$$v ::= pval \mid x \mid \langle D \rangle_{type} \mid \langle D \rangle^{url^{\bot} \rightarrow pat}_{type} \mid x<v> \, \mid ref \mid \ldots $$

$$pval ::= \ok \mid \err \mid num \mid \ldots$$

where $url^{\bot}$ stands for either $url$ or $\bot$. The two cases arise depending if
the deployed component is an application component (in this case the $url$ represents the codebase $url$) or not.

$D$ is a declaration of a set of operations $op$, which are user-defined or commands, defined by

$$op ::= \com \mid x$$

$$\com ::= \pu \mid \get \mid \delete \mid \rexec$$

Formally, $D$ is a partial function from operations $op$ to pairs composed by a formal parameter variable $x$ and a term $E$
representing the code of the operation.
Definitions in $D$ are represented as $op(x) \eqdef E$. Moreover, we assume
$D$ to be such that, if $\com(x) \eqdef E \in D$, for $\com \in \{\get, \delete\}$ , then $x \notin fr(E)$~\footnote{We use $fr(E)$ to denote free names included in $E$.}, i.e. no value is received by the $\get$ and $\delete$ operation definitions. In general, in the following, we will just use $op \eqdef E$, to stand for a definition $op(x) \eqdef E$ such that $x \notin fr(E)$.

We are now in a position to represent the syntax of terms $E$, i.e. code defining operations. We preliminary need some definitions

In method code we use urls starting with 
$<\session>$, $<\application>$  and $<\phbase>$ to manage session attributes of the current application and application attributes
(for service components) and access the 
physical base (for application components).

We also use relative pathes starting with 
$<\ipath>$ to access the internal path, i.e. the URL suffix identifying the resource on which a service component is called.

References $ref$ are possible way of expressing addresses (relative, root-relative or absolute) in a middleware command.

$$ref ::= url_s \mid rpath_s \mid \bs rpath \mid \bs \exec \bs rpath$$

where $url_s$ is defined as for $url$ except that $burl_s$ replaces $burl$
and $rpath_s$ is defined as for $rpath$ except that $bpath_s$ replaces $bpath$.
The syntax of $burl_s$ is as that of $burl$ with the additional production
$$burl_s ::= \dots \mid <symurl>$$
where
$$symurl ::= \session \mid \application \mid \phbase \mid x$$
The syntax of $bpath_s$ is as that of $bpath$ with the additional production
$$bpath_s ::= \dots \mid <\ipath>$$.

We also need to introduce expressions. An expression $e$ 
includes $v$ possibly combined with operators and returns a $v$.
A boolean expression $be$ includes $v$ possibly combined with operators and returns a boolean ($true$ or $false$).
In the semantics we will denote evaluation of expressions $e/be$ such that all variables (unbound names) have been already instantiated with $\cale(e)/\cale(be)$.

Finally, we need to introduce session delegation sets $rs$, which
specify a set of contexts for which we want the client session
to be delegated. $\varepsilon$ included in the list is a relative reference to the context of the base URL (codebase for application components, physical base for service components).

$\begin{array}{l}
rs ::= \{ rlist \} \, \mid \{\} \mid \cali \\
rlist ::= \lo , rlist \mid \lo \mid \varepsilon
\end{array}$

Notice that the spacial case $\cali$ of $rs$ denotes an internal command, i.e. a usage of a default command behaviour inside
a user defined command/operation. In this case the command expects
a relative url identifying the resource it must work with.

$\begin{array}{ll}
E ::= 
& x = {\bf put}^{rs}_{ref} \, e.E \; |\\
& x = {\bf get}^{rs}_{ref}.E \; |\\
& x = {\bf delete}^{rs}_{ref}.E \; |\\
& x = {\bf rexec}^{rs}_{ref} \, e.E \; |\\
& x = {\bf lexec}^{rs}_{ref} \, e.E \; |\\
& x = e.E \; |\\
& \overline{x} \; e.E |\\
& x(y).E |\\
& {\bf spawn} \;E\,.E | \\
& {\bf if} \, be \;\, {\bf then} \, E \; {\bf else} \, E \; | \\
& \nu \!<\!session\!\!>.E \; | \\
& \neg \!<\!session\!\!>.E \; | \\
&{\bf return} \; e \\
& \nil
\end{array}$ \\
where, for commands $x = \com^{rs}_{ref} e.E$ occurring in $E$, we have that $rs=\cali$
implies $\exists rpath_s: ref = rpath_s$.

\subsection{Semantics}

In order to present the semantics we need to preliminary define
the syntax of terms $P$ representing the syntax of programs/operations in execution.

In order to do this we need to extend the syntax of urls
(terms $burl$), as defined in the previous section, so to represent session managing via special resources.

$$burl ::= \dots \mid \lo \bs extrapath \mid x$$

$$extrapath ::= \session \mid \session \bs \ns \mid \application$$

$S$ is a metavariable that can stand for a session identifier (a name $x$ or no session $ns$, in the case no session is detained).

$$\ses ::= \ns | x$$

We need also to introduce session delegation pairs $sls$ that are transmitted inside a service request.

$\begin{array}{l}
sls ::= \{ sllist \} \, \mid \{\} \mid \cali \\
sllist ::= \lo\!:\!\ses , sllist \mid \lo\!:\!\ses
\end{array}$

The syntax of terms $P$ representing operations in execution is as follows.

$\begin{array}{ll}

P ::= 
& x = {\bf put}^{sls}_{url:\ses} \, e.P \; |\\
& x = {\bf get}^{sls}_{url:\ses}.P \; |\\
& x = {\bf delete}^{sls}_{url:\ses}.P \; |\\
& x = {\bf rexec}^{sls}_{url:\ses} \, e.P \; |\\
& x = {\bf lexec}^{sls}_{url:\ses} \, e.P \; |\\
& x = \, e.P \; | \\
& {\overline{x}\,}^{sls} \; e.P |\\
& x(y).P |\\
& (\nu \, x) P \; | \\
& {\bf spawn} \;P\,.P | \\
& {\bf if} \, be \;\, {\bf then} \, P \; {\bf else} \, P \; | \\
& \nu \; \lo \bs \session \bs \ses . P \; | \\
& \neg \; \lo \bs \session \bs \ses  . P \; | \\
& \nil
\end{array}$

Before presenting the semantics, we need to introduce pathes $path$ as the path information 
that can occur in an url after the context.

$$path ::=  \bs rpath \mid  \bs \exec \bs rpath \mid \bs extrapath \bs rpath \mid \varepsilon$$

In the following we will use $\calr$ (resource collections)
to denote terms $N$ that do not include $(\nu n) N$ subterms.

The semantics is defined in tables~\ref{congrules},~\ref{basicrules},~\ref{uncaptcom},~\ref{captcom},~\ref{sesrules} and~\ref{lexecrules}.
In the tables we assume the following definitions.

\begin{itemize}
\item
$Int_g=\{ \lo \bs,\lo \bs \session \bs, \lo \bs \exec \bs \mid \lo \in \call \}$
\item
$Int_d=\{ \lo \bs,\lo \bs \application \bs \mid \lo \in \call \}$
\item

$maxpat(\calr,path)= max \;
\{pat \in pats(\calr) \mid path \in pat \}$~\footnote{
We assume that $\varepsilon$ is returned (the smallest element in the prefix relation) if we have an emptyest.}\\
where $pats(\calr) = \bigcup_{[R]_{url} \in \calr } pat([R]_{url})$.
$pat([R]_{url}) = \{pat\}$ if $R = \langle D \rangle^{cbase \rightarrow pat}_{type}$ and
$url = \lo \bs \exec \bs m \bs$ for any $D$, $cbase$, $pat$, $\lo$ and $m$; 
$pat([R]_{url}) = \emptyset$ otherwise.

\item $match(\lo \, path,pat,\calr) = \lo \, path \in urls(\calr) \wedge path \in pat \wedge pat \geq maxpat(\calr_l, path)$
\end{itemize}

Moreover $d(pat)$ returns the path corresponding to the directory
of the $pat$ (after that $*$ has been removed), $\hookrightarrow$ returns the argument on the left if it is not $\bot$, otherwise
it returns the argument to the right, $url(\_,\_)$ and $l(\_,\_)$
perform the expected evaluation of an (absolute) $url$ and a
context $\lo$ starting from the absolute base $url$ on the left and the relative url
on the right. $path-pat$ computes the $path$ suffix matching $*$ (or $\varepsilon$ if there is no $*$) in the expected way.

Finally, the boolean function $cond$ is assumed to be any predefined function such that, for any context $\lo$:
$cond(\lo\bs \session \bs) =
cond(\lo\bs \exec \bs) =
cond(\lo\bs) = true$.
Moreover, $loc_{l, type}$ is assumed to be a predefined
function that returns a context $l'$ located at the same IP (in the
same machine) as $l$, such that $l'$ is capable to execute the client-side technology of type $type$. $type(x)$ determines
the type of value $x$: recall that values represent (content of)
typed files. Hence we have $type(\langle D \rangle_{type})=type$.

Notice that the negative premise in Table~\ref{basicrules} does not cause bad definedness of
the operational semantics. This because,
in the rules for captured commands, the capability of a term of performing some auxiliary transition is {\bf not} conditioned on the existence of some reduction transition $\arrow$ (which could in turn depend on auxiliary transitions), this
because the term $N \pa \calr$ considered in the premise can always perform a reduction transition (i.e. the synchronization between $\overline{op}$ and $op$).

\begin{table}[htbp]
\caption{Congruence rules.}
\label{congrules}
\centering
$\begin{array}{rcll}
\hline\\[-.2cm]
N_1 \pa N_2 & \equiv & N_2 \pa N_1 &\\[.2cm]
(N_1 \pa N_2) \pa N_3 & \equiv & N_1 \pa (N_2 \pa N_3) &\\[.2cm]
((\nu x) N_1) \pa N_2 & \equiv & (\nu x) (N_1 \pa N_2) & x \notin fr(N_2)\\[.2cm]
[(\nu x) P]_{url} & \equiv & (\nu x) [P]_{url} & \\[.1cm]
\hline
\end{array}$
\end{table}

\begin{table}[htbp]
\caption{Basic rules.}
\label{basicrules}
\centering
$
\begin{array}{c}

\hline
\infr{N \arrow N' \wedge N_1 \equiv N}
{N_1 \arrow N'}
\hspace{1cm}\infr{N \arrow N'}
{(\nu x) N \arrow N'}
\\[.5cm]
\infr{[x = \com^{sls}_{url:\ses} \hat{e}.P]_{url'} \pa \calr \parrow{} \calr'}
{[x = \com^{sls}_{url:\ses} \hat{e}.P]_{url'} \pa \calr \arrow{} \calr'} \\[.5cm]
\infr{[x = \com^{sls}_{url:\ses} \hat{e}.P]_{url'} \pa \calr \nparrow{}}
{[x = \com^{sls}_{url:\ses} \hat{e}.P]_{url'} \pa \calr \arrow{} [P\{\err/x\}]_{url'} \pa \calr}
\\[.5cm]
\begin{array}{ll}
[\spawn \, P.Q]_{burl\bs} \pa \calr \arrow{} & t \notin fr(P,burl)\\[.1cm]
\hspace{1.8cm} [Q]_{burl\bs} \pa ((\nu t) [P]_{burl \bs t \bs}) \pa \calr & \\[.3cm]
[\myif be \mythen P \myelse Q]_{url} \pa \calr \arrow{} [P]_{url} \pa \calr & \cale(be)= true
\\[.3cm]
[\myif be \mythen P \myelse Q]_{url} \pa \calr \arrow{} [Q]_{url} \pa \calr & \cale(be)= false
\\[.3cm]
\end{array} \\[0cm]
\begin{array}{l}
[\overline{op}^{\, sls} e.P]_{url} \pa [op(x).Q]_{url'} \pa \calr \arrow{} \\[.1cm]
\hspace{3.7cm}[P]_{url} \pa [Q\{\cale(e)/x\}\theta_1\theta_2]_{url'} \pa \calr  
\\[.2cm]
\theta_1= \{\lo \; path : \ses' \,/\, \lo \; path : \ses \mid path \in 
Path 
\wedge
\; \lo:\ses' \in sls \} \\[.1cm]
\theta_2= \{\lo \bs \session \bs \ses'  \; / \; \lo \bs \session \bs \ses \mid 
\; \lo:\ses' \in sls \}\\[.1cm]
\end{array} \\[.1cm]
\hline
\end{array}$
\end{table}

\begin{table}[htbp]
\caption{Rules for auxiliary transitions of uncaptured commands.}
\label{uncaptcom}
\centering
$\begin{array}{ll}
\hline\\[-.2cm]
[x = \pu^{sls}_{url:\ses} e.P]_{url'} \pa [v']_{url} \pa \calr  \\[.1cm]
\parrow [P\{\ok/x\}]_{url'} \pa [\cale(e)]_{url} \pa \calr&
\\[.1cm]
[x = \pu^{sls}_{url:\ses} e.P]_{url'} \pa \calr & d(url) \in urls(\calr) \cup Int_d \; \wedge \\[.1cm] 
\parrow [P\{\ok/x\}]_{url'} \pa [\cale(e)]_{url} \pa \calr
& id(url) \not\in id(urls(\calr)) \\[.1cm]
[x = \get^{sls}_{url:\ses}.P]_{url'} \pa [v]_{url} \pa \calr  \\[.1cm]
\parrow [P\{v/x\}]_{url'} \pa [v]_{url} \pa \calr& \\[.1cm]
[x = \delete^{sls}_{url:\ses}.P]_{url'} \pa [v]_{url} \pa \calr & \not\exists url'' \in urls(\calr): url'' > url \\[.1cm]
\parrow [P\{\ok/x\}]_{url'} \pa \calr 
\\[.1cm]
[x = \rexec^{sls}_{burl\bs:\ses} e.P]_{url'} \pa \calr 
& burl \bs \in urls(\calr) \cup Int_g \, \wedge \\[.1cm]
\parrow (\nu n)([P\{n/x\}]_{url'} \pa & n \notin fr(\{P,e,burl,url'\})\\ [.1cm]
\hspace{.6cm} [\cale(e)]_{burl\bs n \{\bs\}^{cond(burl \bs)}})  \pa \calr &  \\[.5cm]
\multicolumn{2}{l}{
\begin{array}{l}
\mbox{additional condition for each rule:} \\[.2cm]
sls \, = \cali\; \,
\vee \not\exists r \in \calr: pat(r)=
\{ maxpat(\calr, url)\} \wedge \com \in ops(r) 
\\[.2cm]
\mbox{where} \; \com \; \mbox{is the rule command and} \; url=burl\bs \; \mbox{for the} \; \rexec \; \mbox{rule}
\\[.1cm]
\end{array}
}\\[.1cm]
\hline
\end{array}$
\end{table}

\begin{table*}[htbp]
\caption{Rules for auxiliary transitions of captured commands.}
\label{captcom}
\centering
$\begin{array}{l}
\hline
\infr
{
N \pa \calr \arrow \calr' \wedge op(y) \eqdef Q \in D \wedge match(\lo \, path,pat,\calr)
} 
{
[x = op^{sls}_{\lo path:S} \, e.P]_{url'} \pa [\langle D \rangle^{ cbase \rightarrow pat}_{type}]_{\lo \bs \exec \bs m \bs} \pa \calr \parrow \calr'
}
\hspace{.3cm}
op \in Com
\\[.6cm]
\infr
{
N \pa \calr \arrow \calr' \wedge op(y) \eqdef Q \in D \wedge match(\lo \, path,pat,\calr)
}
{[x = \rexec^{sls}_{\lo path: \ses} \, op<e>.P]_{url'} \pa [\langle D \rangle^{cbase \rightarrow pat}_{type}]_{\lo \bs \exec \bs m \bs} \pa \calr \parrow \calr'
} 
\hspace{.3cm}
\begin{array}{l}
op \notin Com \, \wedge \\[-.2cm]
\rexec \notin dom(D) \, \wedge \\[-.2cm]
path = path'\bs n
\end{array}
\\[.7cm]
N \; = \; ((\nu z)([\overline{op}^{\, sls \cup \{\lo : \ses \}} e.z(x).P]_{url'} 
\pa 
\\[0pt]
(\nu t)[op(y).Q\theta_1\theta_2\{\bar{z}^{sls \cup \{\lo : \ses \}}/return\}]_{\lo \bs \exec \bs m \bs t \bs}))
 \pa [\langle D \rangle^{cbase \rightarrow pat}_{type}]_{\lo \bs \exec \bs m \bs}
 \\[.2cm]
\theta_1=\{ \lo \bs \session \bs \ns \; / <session> \}  \{ \lo \bs \application \; / <application> \} 
\\
\{ \lo \; id(d(pat)) \; / <phbase>  \}  \{ path-pat \; / <ipath> \} \\[.2cm] \theta_2=
\{ x=\com^{\cali}_{url(pat,rpath):\ns} e.P / x=\com^{\cali}_{rpath} e.P \mid x=\com^{\cali}_{rpath} e.P \in \calp \} \\[.2cm] 
\tabspace{\theta_2=\;}\{ x=\com^{\{l(cbase \hookrightarrow \lo,r):\ns \mid r \in rs \}
}_{url(cbase \hookrightarrow \lo \, d(pat),ref):\ns} e.P / x=\com^{rs}_{ref} e.P \mid x=\com^{rs}_{ref} e.P \in \calp \}\\[.2cm]
\hline
\end{array}
$

\end{table*}

\begin{table}[htbp]
\caption{Rules for sessions.}
\label{sesrules}
\centering
$\begin{array}{ll}
\hline \\[-.1cm]
[\nu \, \lo\bs \session \bs \ses . P]_{url} \pa \calr \arrow [P]_{url} \pa \calr & \ses \neq \ns \\[.2cm]
\infr
{[x = \rexec_{\lo\bs \session \bs: \ns }. P\theta_1\theta_2]_{url} \pa \calr \arrow \calr'}
{[\nu \, \lo\bs \session \bs \ns . P]_{url} \pa \calr \arrow \calr'} \hspace{.1cm} & 
x \notin fr(P)
\\[.4cm]
\multicolumn{2}{l}{
\theta_1=\{\lo\bs \session\bs x  \; / \; \lo\bs \session\bs \ses \, \mid  \ses \in \caln \cup \{ \ns \} \}
}\\[.1cm]
\multicolumn{2}{l}{
\theta_2=\{\lo \; path : x \; / \; \lo \; path : \ses \mid path \in Path
\wedge \ses \in \caln \cup \{ \ns \} \}
}
\\[.3cm]
[\neg \, \lo\bs \session \bs \ns. P]_{url} \pa \calr \arrow [P]_{url} \pa \calr & 
\\[.2cm]
\infr 
{[\delete_{\lo\bs \session\bs \ses \bs : \ns}. P\theta_1\theta_2]_{url}\pa \calr \arrow \calr'}
{[\neg \, \lo\bs \session \bs \ses. P]_{url} \pa \calr \arrow \calr'} \hspace{.1cm} & \ses \neq \ns\\[.4cm]
\multicolumn{2}{l}{
\theta_1=\{\lo\bs \session\bs \ns \; / \; \lo\bs \session\bs \ses \mid 
\ses \in \caln \cup \{ \ns \} \}
}\\[.1cm]
\multicolumn{2}{l}{
\theta_2=\{\lo \; path : \ns \; / \; \lo \; path : \ses \mid path \in Path
\wedge \ses \in \caln \cup \{ \ns \} \}
}\\[.1cm]
\hline
\end{array}$
\end{table}

\begin{table}[htbp]
\caption{Rules for local execution.}
\label{lexecrules}
\centering
$\begin{array}{ll}
\hline\\[-.2cm]
\infr
{
\begin{array}{l}
[x = \get_{\lo \, path \bs n \,:\ses}. \lop= loc_{\lo,type(x)} . \, 
\\
n' = \rexec_{\lop \bs\,:\ns}. 
\rexec_{\lop\bs \exec \bs :\ns} x^{\lo \, path \bs \rightarrow \lop \bs n' \bs n}. 
\\
\pu^{sls \cup \{\lo : \ses \}}_{\lop \bs n' \bs n \,:\ns} e.P\{\lop\bs n' \, /  y\}]_{url}
\pa \calr 
\arrow \calr'
\end{array}
}
{[y = \lexec^{sls}_{\lo \, path \bs n :\ses} e.P]_{url} \pa \calr \arrow \calr'} \\[.9cm]
\hline
\end{array}$
\end{table}

\section{Conclusion}

We experimented a Java based partial implementation of the middleware, implementing our integrated version of interface-based and restful based web services, with comet related applications (application that send events in real time to the front-end interface) and we tested several solutions base on services keeping responses permanently open and long polling solutions. Moreover we experimented with the performance of several data binding methods.

\bibliographystyle{plain}

\end{document}